\documentclass[pra,twocolumn,showpacs,groupedaddress,nofootinbib]{revtex4-1}
\usepackage{amsmath,amssymb,amsfonts}
\usepackage{graphicx}
\usepackage{times}

\usepackage[unicode,breaklinks]{hyperref}
\hypersetup{
    unicode=true,
    a4paper=true,
    plainpages=false,
    pdftitle={Quantum reflection of bright solitary matter-waves from a narrow attractive potential},
    pdfauthor={},
    pdfsubject={Quantum reflection of bright solitary matter-waves from a narrow attractive potential},
    colorlinks=true,
    linkcolor=blue,
    citecolor=blue,
    filecolor=black,
    urlcolor=blue
}
\begin{document}


\title{Production and characterization of a dual species magneto-optical trap of cesium and ytterbium} 



\author{S. L. Kemp$^{1}$, K. L. Butler$^{1}$, R. Freytag$^{1,2}$, S. A. Hopkins$^{1}$, E. A. Hinds$^{2}$, M. R. Tarbutt$^{2}$, and S. L. Cornish$^{1}$}

\affiliation{$^{1}$Joint Quantum Centre (JQC) Durham-Newcastle, Department of Physics, Durham University, South Road, Durham, DH1 3LE, United Kingdom}

\affiliation{$^{2}$Centre for Cold Matter, Blackett Laboratory, Imperial College London, Prince Consort
Road, London, SW7 2AZ, United Kingdom}

\date{\today}


\date{\today}

\begin{abstract}
We describe an apparatus designed to trap and cool a Yb and Cs mixture. The apparatus consists of a dual species effusive oven source, dual species Zeeman slower, magneto-optical traps in a single ultra-high vacuum science chamber, and the associated laser systems. The dual species Zeeman slower is used to load sequentially the two species into their respective traps. Its design is flexible and may be adapted for other experiments with different mixtures of atomic species. The apparatus provides excellent optical access and can apply large magnetic bias fields to the trapped atoms.  The apparatus regularly produces $10^{8}$ Cs atoms at 13.3~$\mu$K in an optical molasses, and $10^{9}$ Yb atoms cooled to 22~$\mu$K in a narrowband magneto-optical trap.  
\end{abstract}


\maketitle 



\section{Introduction}

The production of ultracold atomic mixtures has opened the door to a wide range of interesting quantum physics~\cite{Greiner2002,Krinner2015,Khaykovich2002,Matthews1999a}. Moreover, the ability to produce ultracold heteronuclear polar molecules from quantum gases of their constituent atoms leads to many potential applications~\cite{Carr2009}. These molecules typically have large ground state permanent dipole moments giving rise to long-range anisotropic interactions. Such molecules may be used for studying ultracold chemistry~\cite{Krems2008,Ospelkaus2010,DeMiranda2011}, novel phase transitions~\cite{Wall2009,Micheli2007,Capogrosso-Sansone2010,Buechler2007}, and opening new paths for precision measurement~\cite{Hudson2011,Flambaum2007,Isaev2010,Baron2014}. Due to their long-range dipole-dipole interactions, polar molecules can also be used to simulate the behavior of strongly-interacting many-body quantum systems which are too complex to model on a computer \cite{Feynman1982}. A lattice of ultracold polar molecules can simulate a range of Hamiltonians~\cite{Georgescu2014,Lewenstein2007} or act as a processor of quantum information~\cite{Micheli2006,DeMille2002,Baranov2012}.

Whilst several groups have created ultracold ground-state  bi-alkali dimers~\cite{Danzl2010,Ni2008,Takekoshi2014,Molony2014,Park2015}, a few are now starting to study alkali-alkaline earth systems~\cite{Hara2013,Khramov2014,Borkowski2013,Pasquiou2013}. The unpaired electron spin in such systems gives rise to a permanent magnetic dipole moment in addition to the ground state electric dipole moment, enabling extra control over the molecular interactions and allowing spin lattice models to be simulated~\cite{Micheli2006}. The expanding list of species combinations being used has led to the development of novel cooling techniques and to the design of apparatus for cooling and trapping more than one species~\cite{Stellmer2013}.

In this paper we describe the production and characterization of a dual species magneto-optical trap (MOT) of cesium (Cs) and ytterbium (Yb). We also outline the apparatus designed to study the interspecies scattering length. The apparatus includes a versatile dual species Zeeman slower, which could also be used to decelerate other combinations of species. Yb has seven stable isotopes which, when paired with the high mass of the Cs atom, provide a useful tunability of the reduced mass of the system~\cite{Zuchowski2010a}. This enhances the likelihood of a favorable background scattering length and of finding magnetic Feshbach resonances that arise from the dependence of hyperfine coupling on the internuclear separation~\cite{Brue2013,Brue2012}. In order to calculate where Feshbach resonances lie, it is important to know the background interspecies scattering length of the system for one isotopic combination. This may be achieved by measuring the energy of the topmost molecular bound state using 2-photon photoassociation spectroscopy~\cite{Kitagawa2008,Abraham1997,Jones2006,Muenchow2011}, or by rethermalisation measurements~\cite{Anderlini2005,Bloch2001}. The dual species MOT we describe here is the first step towards such measurements, and the apparatus we detail in this paper is designed with both of these experiments in mind. With the scattering length measured for one isotopologue, mass scaling may be applied to predict Feshbach resonances in all isotopologues, which may then be used to form loosely-bound ultracold molecules~\cite{Chin2010,Koeppinger2014}; the first step towards ground state CsYb molecules.

The structure of the paper is as follows. We first provide an overview of the whole system before describing each of the main components: a dual species atomic oven, a dual species Zeeman slower, a vacuum chamber for the magneto-optical traps (MOTs), and laser systems for each of Cs and Yb. Then we provide details of the performance of both MOTs, including measurements of atom number and temperatures, before closing with a conclusion and an outlook.


\section{System Overview}\label{sec:SystemOverview}
\begin{figure*}
	\centering
		\includegraphics[width=1\linewidth]{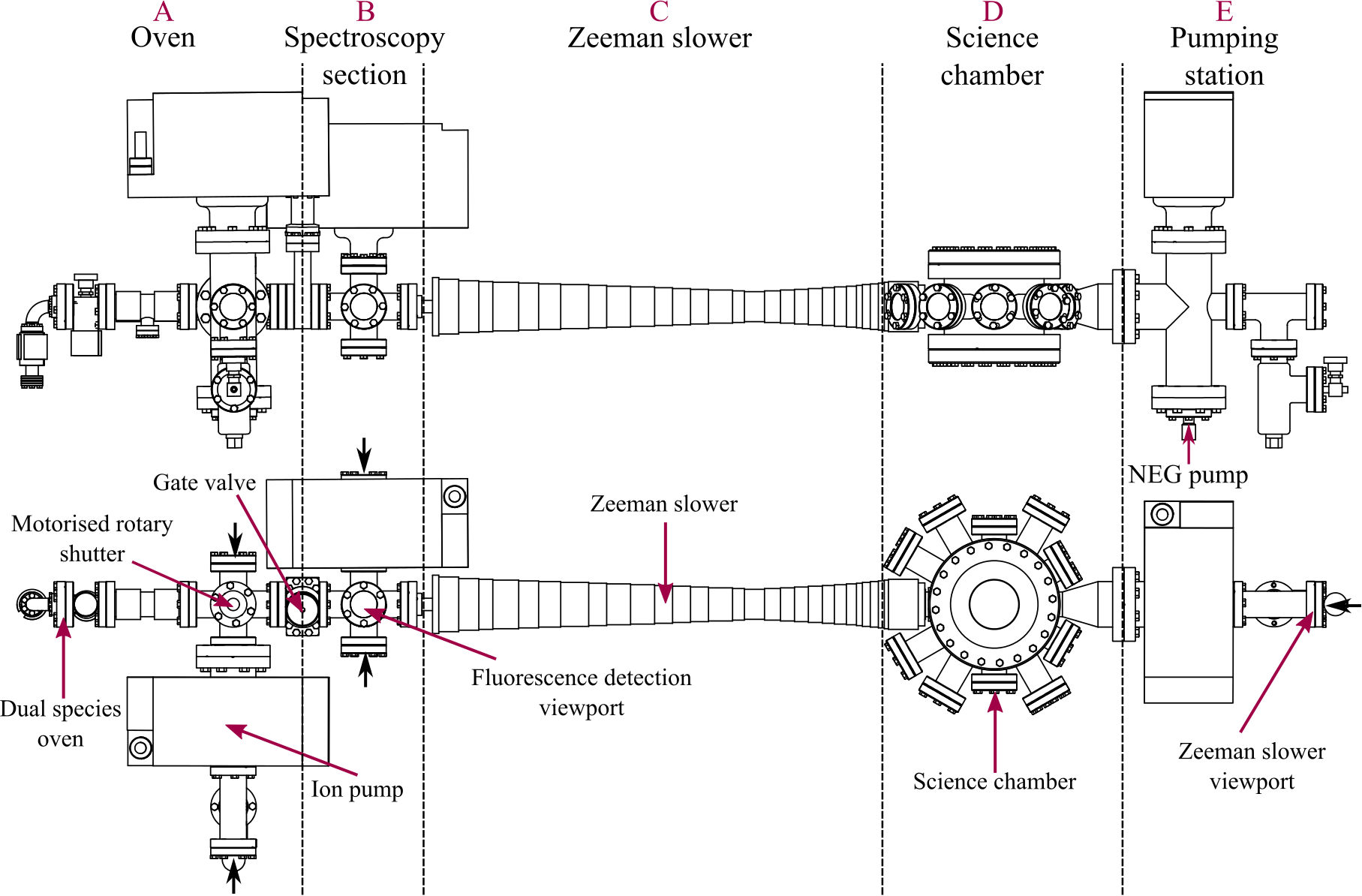}
	\caption{Side and top schematics of the vacuum system with each of the five main sections labeled: A, the dual species oven; B, the spectroscopy section; C, the Zeeman slower; D, the science chamber; and E, its associated pumping station. Several features of the system are highlighted by the red arrows. Black arrows indicate optical access in addition to that present on the science chamber.}
	\label{fig:vacuum}
\end{figure*}

The primary scientific aims of this experiment are to capture both Cs and Yb atoms in a dual species MOT, to transfer these atoms into an optical trap for further evaporative cooling and to find a route to the production of ultracold polar molecules using the techniques of photoassociation or magnetoassociation. These goals impose several critical design specifications on the apparatus: good optical access for the MOT, imaging and optical trapping beams, an ultra-high vacuum (UHV) environment to ensure long trap lifetimes, the creation of large magnetic bias fields in order to access interspecies Feshbach resonances, and the capacity to add an optical lattice to stabilize CsYb against the exchange interaction,  2CsYb$\rightarrow$Cs$_{2}+$Yb$_{2}$~\cite{Zuchowski2010}. 

To meet the design specifications we choose to load atoms into a dual species MOT in a single stainless steel science chamber using a dual species Zeeman slower. This UHV chamber features twelve anti-reflection coated viewports, including two mounted in re-entrant flanges. The purpose of these flanges is to allow magnetic field coils to be mounted close to the atoms, generating large fields without compromising optical access. A further two access ports on the chamber are reserved for Zeeman slowing and a pumping station. Low pressures are maintained throughout the apparatus by a total of three ion pumps and two non-evaporable getter (NEG) pumps as seen in figures~\ref{fig:vacuum} and~\ref{fig:cross}.

The dual species Zeeman slower is designed for sequential loading of the two species into their respective traps, decelerating atoms from an effusive dual species oven. The Zeeman slower needs careful design to load an Yb MOT directly on the narrow 555.8~nm $^{1}S_{0} \rightarrow {}^{3}P_{1}$ transition. This transition has the advantage of a low Doppler temperature but has a correspondingly lower capture velocity (see table~\ref{tab:properties} which summarizes the key properties of Yb and Cs relevant to the design of the apparatus). Direct loading into a 555.8~nm MOT circumvents the need for a collection MOT using the 398.9~nm $^{1}S_{0} \rightarrow {}^{1}P_{1}$ transition, thus saving optical access. It also avoids the problem of optical pumping into the metastable $^{3}D$ levels, which limits the lifetime of a 399~nm MOT~\cite{Honda1999}.

The design and performance of these critical elements of the apparatus are discussed in more detail in the following sections. The result of all these considerations is an elegant dual species apparatus capable of producing high numbers of trapped Cs and Yb atoms.

\begin{table*}[ht]%

\caption{A table summarizing the key properties of Cs and Yb and the associated transitions used in this apparatus.}
\begin{center}
\begin{tabular}{| c | c | c |c|}

\hline
\hline
 & Cs & \multicolumn{2}{c|}{Yb} \\ \hline
 Number of stable isotopes & 1 & \multicolumn{2}{c|}{7} \\
 Vapor pressure, $P_{\rm{V}}$ ($T=300$~K)~\cite{Steck2010b, Alcock1984} (torr) & 7$\times 10^{-7}$ & \multicolumn{2}{c|}{3$\times 10^{-21}$} \\

Temperature at which $P_{\rm{V}}=10^{-3}$~torr ($^{\circ}$C) & 109 & \multicolumn{2}{c|}{408} \\ \hline
Transition & $6$\,$^{2}$S$_{1/2}\rightarrow6$\,$^{2}$P$_{3/2}$~~& 6s$^{2}$\,$^{1}$S$_{0}\rightarrow$\,6s\,6p\,$^{1}$P$_{1}$~~~&  6s$^{2}$\,$^{1}$S$_{0}\rightarrow$\, 6s\,6p\,$^{3}$P$_{1}$\\
\hline
Wavelength, $\lambda$ (nm) & 852.3 & 398.9 & 555.8 \\
Linewidth, $\Gamma_{0}/2\pi$ (MHz) & 5.234 & 28.0 & 0.182 \\
Doppler Temperature, $T_{\rm{D}}$ ($\mu$K) & 126 & 673 & 4.4\\
Saturation intensity, $I_{\rm{sat}}$ (mW\,cm$^{-2}$) & 1.10 & 63.1 & 0.139 \\
Maximum acceleration, $a_{\rm{max}}$~(m\,s$^{-2}$) & $5.79\times 10^{4}$ & $5.06\times 10^{5}$ & $2.36\times 10^{3}$ \\
\hline
\hline
\end{tabular}
\end{center}
\label{tab:properties}
\end{table*}


\section{Dual species oven}
Table~\ref{tab:properties} shows the vapor pressures for Cs and Yb at room temperature. For Yb it is so low that an effusive oven is needed to create a high-flux, collimated atomic beam, which is brought to rest by a Zeeman slower. This the most common approach to loading an Yb MOT, although other methods of loading have also been demonstrated~\cite{Doerscher2013,Rapol2004}. Whilst a MOT of Cs can be produced directly from a vapor, the  high background pressure is not suitable for making ultracold or quantum degenerate clouds. Therefore Cs should be loaded from either a second collection MOT~\cite{Haendel2012,Harris2008} or a Zeeman slower. We have opted to use a dual species oven and Zeeman slower on our system as the more elegant solution, retaining optical access.

\begin{figure}
	\centering
		\includegraphics[width=1\linewidth]{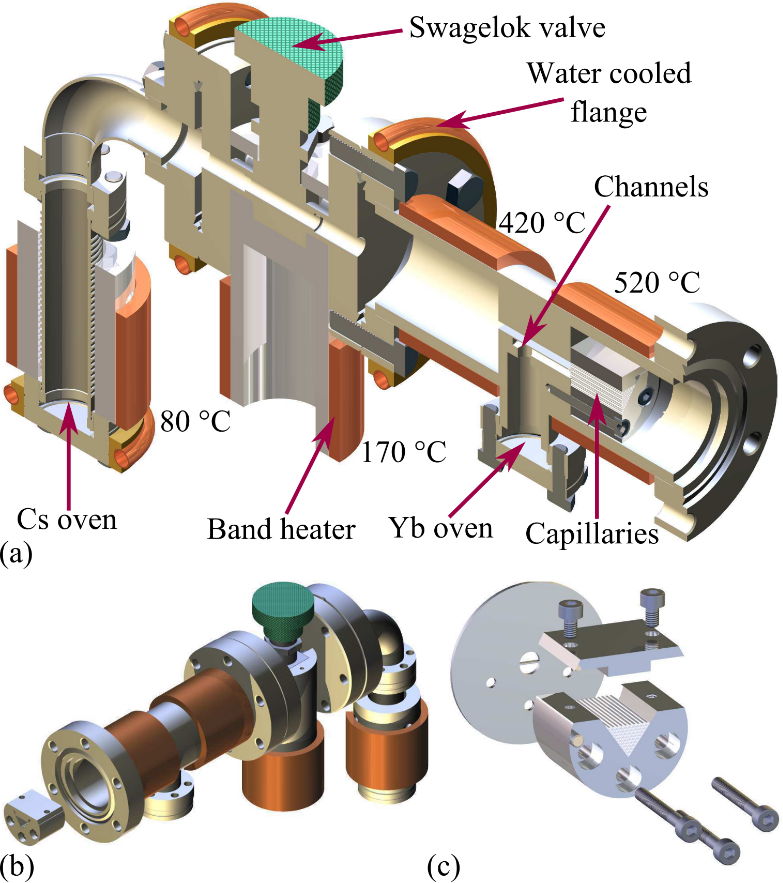}
	\caption{(a) A cutaway schematic of the dual species oven detailing the routes the two species take through the oven. The nozzle heaters are shown as copper bands and the water cooling as copper tubes. The temperatures displayed are the nozzle heater set points, the Yb chamber is typically measured to be at 485\,$^{\circ}$C. (b) Rendering of the oven with the capillary clamp shown outside the oven for detail. (c) Exploded view of the capillary array and the wire-eroded hemispherical channels that the array is clamped over within the oven. Only the face of the custom vacuum part is shown in the figure for clarity.}
	\label{fig:oven}
\end{figure}

Figure \ref{fig:oven} shows a schematic of the dual species oven for Cs and Yb. This design is based on those presented in references \cite{Wille2009,Stan2005}. Two sections separately house the Cs and Yb sources and are heated to different temperatures; the Cs part of the oven to 80\,$^{\circ}$C and the Yb part to 485\,$^{\circ}$C. These numbers are chosen to give approximately the same vapor pressure of $\sim 10^{-3}$~torr for each element. Both sources hang below the axis of the experiment, with the Cs at the far end and the Yb closer to the exit of the oven. The two atomic beams first travel through two channels of a semicircular cross section, which are separated by 0.4 mm. These have been wire-eroded into a custom vacuum part (see figure~\ref{fig:oven} (a) and (c) for the exact geometry).

We require the atomic beam to deliver a high flux to the science chamber located 1.5~m away from the oven. It should be well collimated to avoid Cs and Yb coating the inner surface of the vacuum system in undesired locations. The theory and practice of producing collimated atomic beams has been thoroughly covered in references~\cite{Ross1995,Ramsey1956}. We use an array of 55 capillary tubes, each having an internal diameter of 0.58~mm and length 20~mm. The array is clamped into a triangular recess of angle 70\,$^{\circ}$, which is bolted over the exit of the hemispherical channels (figure~\ref{fig:oven} (b) and (c)).

The Cs part of the oven is loaded using an ampoule containing 1\,g of Cs, placed in the bellows shown at the left hand end of figure~\ref{fig:oven}(a). Once the oven is evacuated and baked, the ampoule is broken, releasing the Cs.  The Yb part of the oven is loaded using a 5\,g ingot of the element. A valve (Swagelok, SS-4H-TW) is used to separate the Cs and Yb, allowing the Cs oven to be sealed when it is not in use. Furthermore, this allows the Yb source to be replenished without bringing the Cs up to air.

The temperature gradients across the oven, shown in figure~\ref{fig:oven}(a), are required to ensure that the sources of Cs and Yb do not migrate from their reservoirs. The capillary tubes need to be the hottest part of the oven in order to avoid the atoms (particularly Yb) sticking to the walls and blocking the tubes. The temperature gradients are provided by a series of four band heaters (Watlow, MB1J1JN2-X66) and water cooled flange jackets, the locations of which are shown in figure~\ref{fig:oven} (a).

\begin{figure}
	\centering
		\includegraphics[width=1\linewidth]{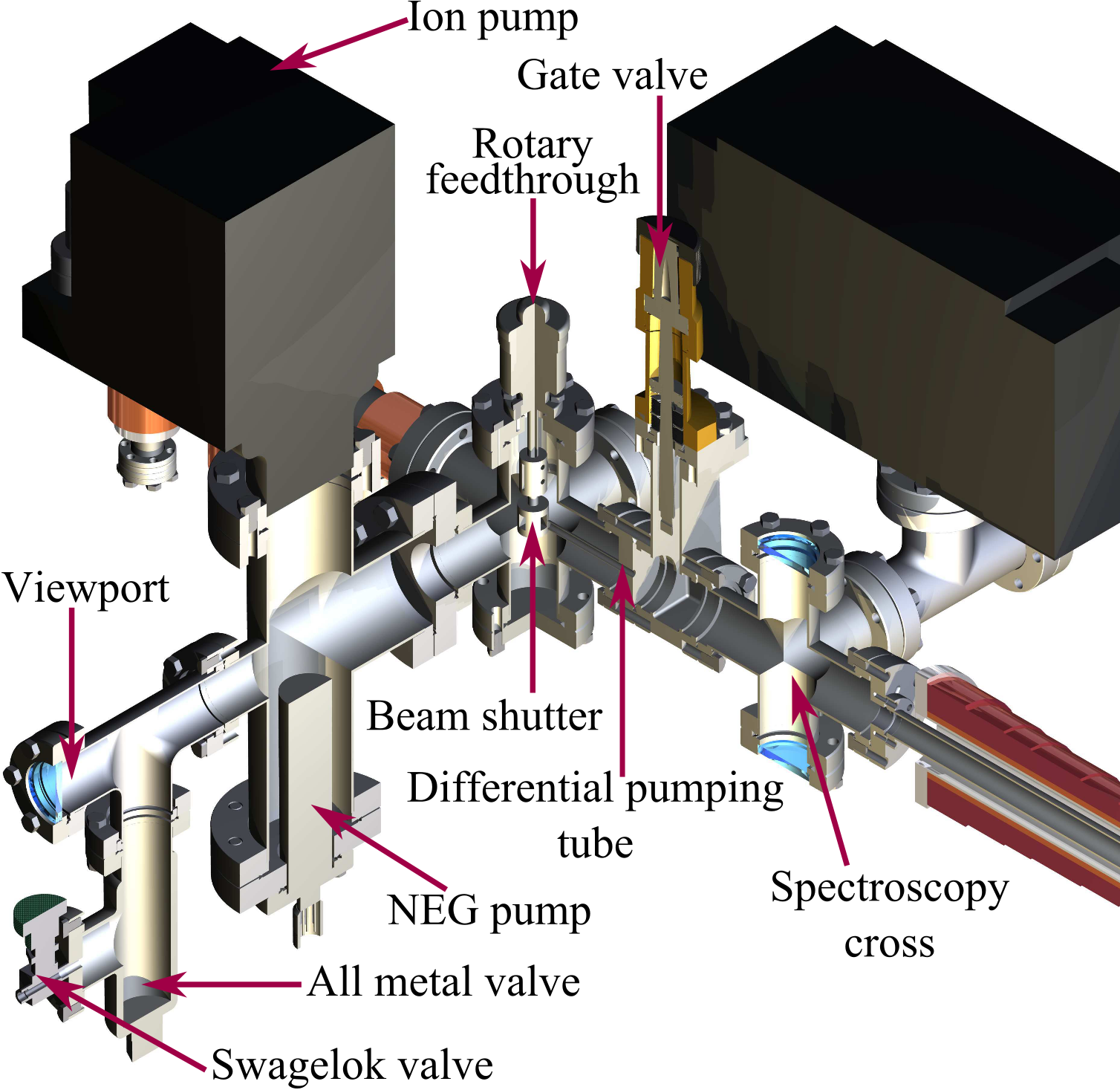}
	\caption{Schematic showing a cutaway of the six-way cross connected to the oven and all of the vacuum pumps, valves, and shutters that are attached to it. Optical access for probing the atomic beam is also shown.}
	\label{fig:cross}
\end{figure}

As seen in figure~\ref{fig:cross}, the oven is connected to a six-way cross, followed by an all-metal DN40 CF gate valve (MDC Vacuum, E-6V-1500M-11). The six-way cross includes connections to an ion pump (Agilent, VacIon55), a NEG pump (SAES, Capacitorr C400-2-DSK), a pair of viewports, and an all-metal valve that allows the oven section to be serviced without letting the science chamber and Zeeman slower up to air. The NEG pump has a large pumping rate for hydrogen within the vacuum system, whereas the ion pump removes any other unwanted atoms or molecules, such as any Cs that builds up over the course of running the experiment. The top flange of the six-way cross is used to connect a rotary motorized feedthrough (MDC Vacuum, BRM-
275-03), that rotates an atomic beam shutter. The shutter features two 5 mm channels orthogonal to one another. This allows the atomic beam flux to be probed when the shutter is open. The rotary motor is controlled by LabVIEW control software allowing the remote blocking of the the beams by rotating the shutter 45\,$^{\circ}$. Two further  water cooling jackets are located on this six-way cross in order to maintain the vacuum system temperature as close as possible to that of the room.


\section{Spectroscopy chamber}

The spectroscopy section ( section B in figure~\ref{fig:vacuum}) is a second six-way cross, with a 40~l\,s$^{-1}$ ion pump, also shown in figure~\ref{fig:cross}, which allows probing of the atomic beams. This is separated from the oven section by a differential pumping tube and the aforementioned all-metal gate valve. The differential pumping tube has a 5~mm diameter and is 60~mm in length, leading to a conductance of 0.25 l\,s$^{-1}$.

Simple absorption spectroscopy was performed on the Cs atomic beam through this six-way cross and at the rotary beam shutter (figure~\ref{fig:cross}) in order to test the collimation of the atomic beam. These measurement points are 23.5~cm apart. A resonant probe beam, orthogonal to the atomic beam, with a $1/e^{2}$ diameter of $0.75\pm0.06$~mm was translated across the atomic beam, and the amount of absorption at each point was recorded. We find the width of the atomic beam to be $6.1\pm0.5$~mm at the rotary shutter and $6.5\pm0.6$~mm at the spectroscopy six-way cross, showing that the oven produces a well-collimated atomic beam. By probing the atomic beam at an angle of 6\,$^{\circ}$ from orthogonal, we observed a Doppler shift of $31\pm1$~MHz of the absorption peak. This corresponds to a mean atomic velocity of $260\pm10$~m\,s$^{-1}$, which is as expected from the Maxwell-Boltzmann distribution for the Cs oven temperature (leftmost band heater in figure~\ref{fig:oven}(a)) at which the data was taken (100\,$^{\circ}$C).


\section{Zeeman slower vacuum}
\label{sec:ZSvac} 
The apparatus incorporates a dual species Zeeman slower, section C in figure~\ref{fig:vacuum}, saving space and optical access to the science chamber by using just one line of entry and one viewport. The vacuum system section of the slower is very simple and consists of a 77~cm long DN16 tube. This has a vacuum conductance of 0.64~l\,s$^{-1}$ and combines with the differential pumping tube to give a pressure  ratio of approximately 10$^{4}$ between the oven section and the science chamber.  The tube is attached to a 3\," flexible coupler (MDC, 075-X), which relieves any strain due to misalignment when the last connection is made between the Zeeman slower and the science chamber. The exit of the Zeeman slower section is 7.5~cm away from the location of the MOT. This distance was minimized in order to prevent slow atoms, in particular the slowed Yb (see section~\ref{sec:ZSdesign}), dropping too far under gravity to be captured in the MOT.


\section{Science chamber}

\begin{figure}
	\centering
		\includegraphics[width=1\linewidth]{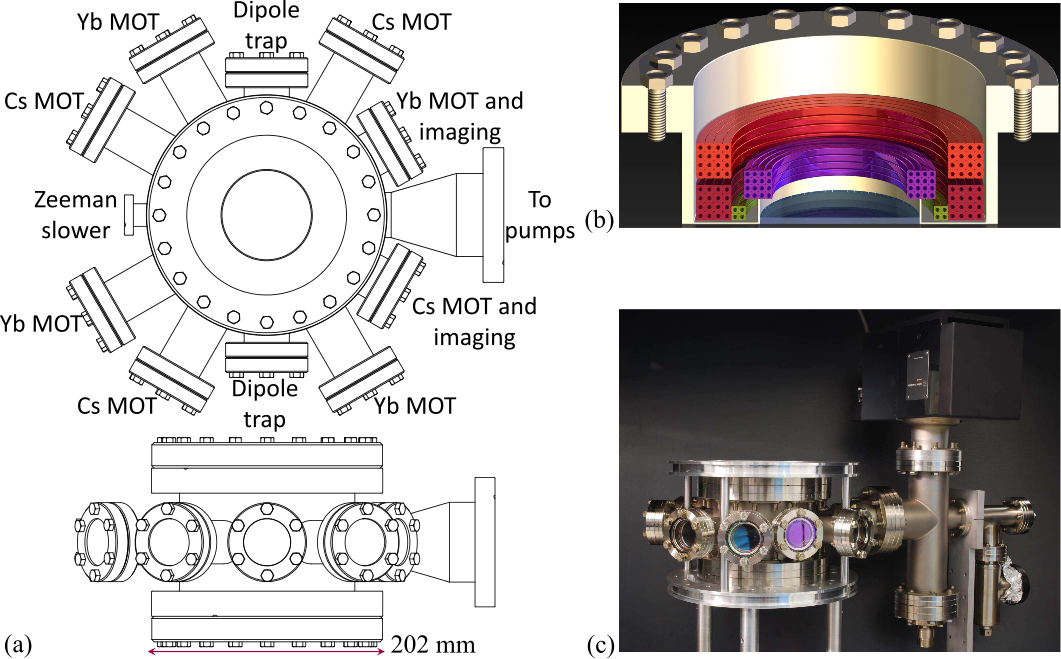}
	\caption{(a) A plan and side view of the science chamber, with each set of viewports labeled.  (b) The top and bottom viewports are set in re-entrant flanges. These are designed to enable bias coils (red, orange and green) to be placed close to the atoms to generate large and uniform bias fields. The purple coils are used for magnetic field gradients. (c) A photograph of the science chamber and pumping station before connection to the rest of the vacuum system.}
	\label{fig:science}
\end{figure}

The science chamber (section D in figure~\ref{fig:vacuum}) is custom made by VG Scienta. It has ten DN40 CF viewports in the horizontal plane for optical access, and two vacuum ports along the beam axis, as shown in figure~\ref{fig:science}(a).

Figure~\ref{fig:science} (b) shows that the top and bottom viewports are set in re-entrant flanges (custom made by the UK Atomic Energy Authority) which allow multiple sets of coil pairs to be mounted as close to the atoms as possible in order to apply magnetic fields up to $\sim$2000~G.  The top and bottom plates of the stand for the science chamber have grooves running around their circumference for `shim' coils to be wound in place which can be seen in the photograph of the finished science chamber, figure~\ref{fig:science}(c). This coil pair produces a bias field of 2.81~G\,A$^{-1}$ in the vertical direction at the center of the chamber, which allows stray magnetic fields to be nulled. Two more pairs of shim coils mounted separately from the science chamber cancel fields in the horizontal plane.

A pumping station is attached to the science chamber to maintain UHV pressures.  This is a twin of the pumping station for the oven.   At the extreme right of figure \ref{fig:vacuum} is a further viewport to allow access for the two overlapping Zeeman slowing laser beams.


\section{Vacuum Assembly}
\label{sec:Vaccum}

The apparatus was assembled in three parts: the oven (figure \ref{fig:vacuum} sections A and B), the Zeeman slower (section C), and the science chamber (sections D and E). Each section was turbo-pumped whilst baking at temperatures up to 200\,$^{\circ}$C in order increase the outgassing rate of the steel. Typically these bakes lasted one week. The sections were then joined together before baking one final time and activating the ion and NEG pumps on the system. Following this procedure the science chamber reached UHV conditions ($<10^{-9}$~torr).

Prior to joining the oven section and the Zeeman slower section, the Zeeman slower magnet was slid over its vacuum tube, which had been wrapped in heater tape. The magnet was supported independently of the vacuum system.  The slower was then attached to the science chamber at one end and the oven section at the other. The whole vacuum system was then pumped into turbo pumps whilst baking on the optical table for two weeks at temperatures up to 200\,$^{\circ}$C. During this period the NEG pumps were activated.


\section{Dual species Zeeman slower}
\label{sec:ZS} 
\subsection{Design}
\label{sec:ZSdesign} 
Several challenges arise in designing a Zeeman slower for two species, which are fully explored in our forthcoming companion paper~\cite{Hopkins2015}. We summarize the important issues now. 
It is worth noting that some pairs of elements can be simultaneously slowed using the same magnetic field profile, depending on atomic parameters such as the wavelength, linewidth and effective magnetic moment of the atomic transition to be used for slowing~\cite{Wille2009}. This does not apply for Cs and Yb, where rather different field profiles are required. Our solution is to design a Zeeman slower that is easily reconfigured and to then slow the two species sequentially by switching the currents in the coil set.  For slowing by scattered laser light there is a maximum theoretical deceleration, 
\begin{equation}a_{\rm{max}}=\frac{\hbar k \Gamma_{0}}{2M},\end{equation} where $k$ is the wavevector of the scattered photons, $\Gamma_{0}/2$ is the saturated scattering rate, and $M$ is the mass of the species being slowed. Typically a parameter $\eta$ is introduced when designing the magnetic field profile of a Zeeman slower, such that a deceleration $\eta a_{\rm{max}}$ is required to maintain slowing. This parameter also determines the minimum length of the slower. This length is $L=v_{\rm{c}}^{2}/\left( 2 \eta a_{\rm{max}}\right)$, where $v_{\rm{c}}$ is the desired capture velocity of the slower. Zeeman slowers are normally designed using $\eta\leq0.6$ to allow some headroom against real world fluctuations away from the ideal configuration such as deviations from the ideal magnetic field profile, photon shot noise and laser intensity variations. For Cs we choose $v_{\rm{c}}=200$~m\,s$^{-1}$ and $\eta_{\rm{Cs}}=0.5$ giving $L=0.7$~m. For Yb we choose $v_{\rm{c}}=300$~m\,s$^{-1}$, and then the same length of slower dictates that $\eta_{\rm{Yb}}=0.128$, a conveniently low value. For both species the required field profiles have the form 

\begin{equation}B(z) = B_{0} + B_{L}\sqrt{1-z/L},   \end{equation} where $B_{L}=hv_{\rm{c}}/\left( \delta\mu \lambda \right)$ and $\delta\mu$ is the difference in magnetic moment between the upper and lower states of the transition. For our design parameters $B_{L}$ is 167~G for Cs and 537~G for Yb.

\begin{figure*}
	\centering
		\includegraphics[width=1\linewidth]{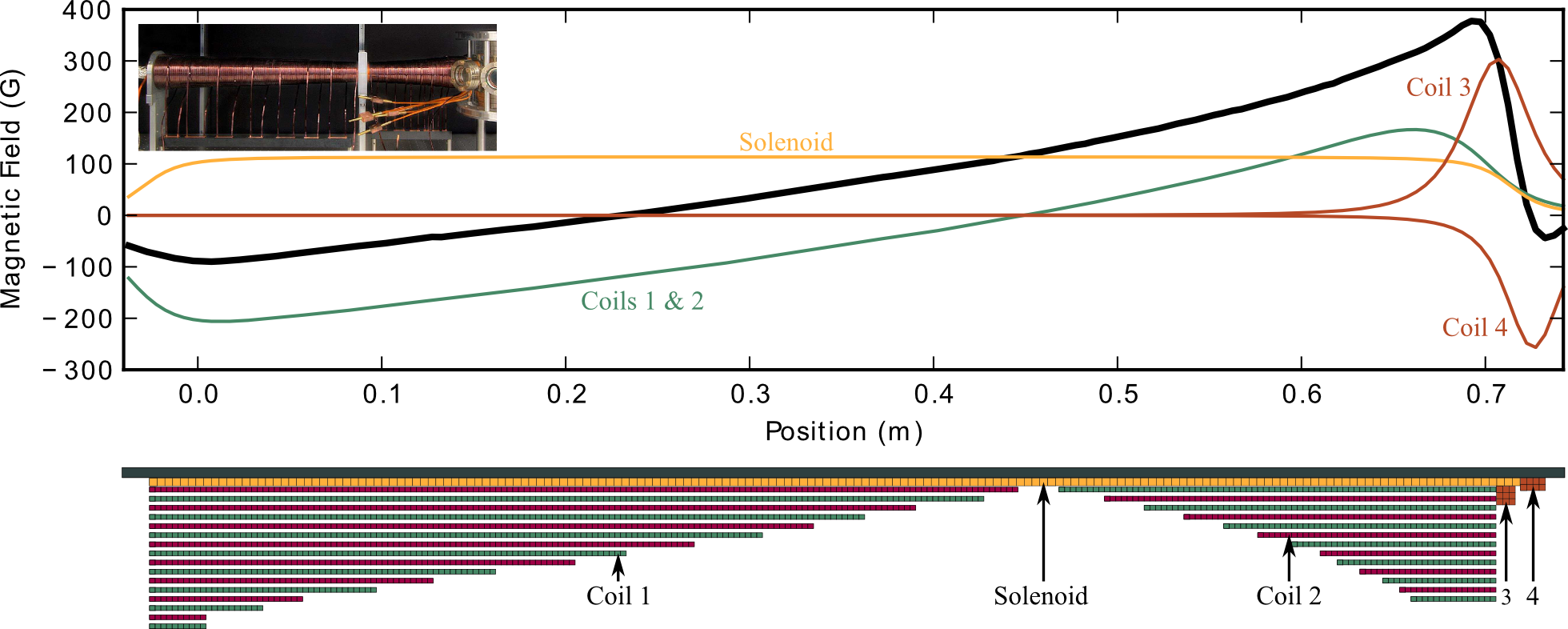}
	\caption{A schematic detailing the Zeeman slower winding pattern, showing the solenoid and the four separate coils that create the field profile. Turns from the 3$\times$1~mm wire are shown in green and red, the 4.3~mm wire turns are shown in yellow and the 3.4~mm wire turns are shown in orange. The solenoid is wound directly onto a steel former tube (grey). The direction of the atomic beam is from left to right. Note that the turns made from the 3$\times$1~mm wire are not to scale and gaps are included between layers for clarity. The turns in red (green) represent windings made from the right (left) to the left (right) of the figure. Above the schematic is a plot showing the measured Zeeman slower field profile when operating at Yb currents (black curve). The contribution to this by each coil is also shown (colored curves matching wire color in schematic). The position scale of this plot matches that of the schematic. The inset shows a photograph of the Zeeman slower showing below the connections between the pairs of layers comprising coils 1 and 2. The high current terminals to the water cooled coils (the solenoid and coils 3 and 4) are also visible to the right of the image.}
	\label{fig:windings}
\end{figure*}

The term $B_{0}$ is the field at the slower exit, where $z = L$, and is a free parameter as far as the slowing process is concerned but must be matched by the laser detuning, $\Delta$, so that $B_{0}\delta\mu = \hbar \Delta$. The choice of detuning is important as the Zeeman slowing beams pass through the MOT region and may perturb the MOTs if the detunings are not appropriate. The final detunings adopted are discussed in section~\ref{sec:lasers}. 

Another important factor for a Zeeman slower is the exit velocity, which must not exceed the capture velocity of the MOT. For Cs the MOT capture speed is around 50~m\,s$^{-1}$ and does not present a great challenge. However, for the 555.8~nm transition in Yb the MOT capture speed is only about 7~m\,s$^{-1}$. Furthermore, the exit speed must be greater than 3~m\,s$^{-1}$ if the atoms are to reach the MOT without falling too far under gravity or diverging excessively. Thus, the slower must deliver atoms in the narrow velocity range of 3 to 7~m\,s$^{-1}$~\cite{Hopkins2015}.

The desired field profile is generated using a solenoid and four separate coils, as shown in figure~\ref{fig:windings}. The solenoid, wound along the entire length of the slower, allows the whole magnetic field profile to be shifted up or down in magnitude to match the slowing light detuning required for Cs or Yb. The direction of current flow in the solenoid can be reversed by a H-bridge switch. The general shape of the field is created by coils 1 and 2; these are run in series but with the current handedness reversed between the two. Coil 3 provides the large field at the end of the slower while coil 4 makes a sharp drop off in field before the MOT region: this is required to minimize the distortion  of the MOT magnetic quadrupole by the Zeeman slower field.

The coil set is designed to give maximum flexibility over the shape of the two magnetic field profiles that are needed. We calculate the fields produced by the coils and then simulate the Zeeman slower to determine the fraction of atoms delivered by the slower in the desired velocity group. We then iterate this calculation to optimize the field profile. The final design profiles for Yb can be seen in figure~\ref{fig:windings}, which shows the separate fields from each coil as well as their sum. The measured profiles for both Yb and Cs are compared in figure~\ref{fig:ZeemanField} with those of our design.


\subsection{Winding and characterization}

The Zeeman slower is wound as follows. First, the solenoid is wound directly onto an 80~cm long steel former tube of outer diameter 38.1~mm. 
The wire is coated with Araldite 2011 as it is wound onto the former. Typically 20 turns are added at a time. Overnight the coil is compressed against the direction of winding and the Araldite is allowed to cure before continuing the winding the following day. The completed solenoid consists of 177 turns of 4.3$\times$4.3~mm  square cross-section wire. This wire is hollow; a 2.75~mm diameter bore runs through the wire allowing cooling water to flow through the coil.

With the former tube still mounted in a lathe, coils 1 and 2 are  wound directly on top of the solenoid, using wire of 3.08$\times$1.19~mm rectangular cross-section. These coils are formed as a series of two-layer coils; each layer pair was wound by starting the first layer from the middle of the slower outwards towards the end of the slower before returning towards the center on the second layer (see figure~\ref{fig:windings}). Once again, these were secured using Araldite 2011. Once a layer pair has cured, the next layer pair is wound directly on top of it until the profile in figure~\ref{fig:windings} is fully built up. The pairs of layers are then connected externally to form a single coil (see inset to figure~\ref{fig:windings}).

Finally, coils 3 and 4 are wound using hollow 3.4$\times$3.4~mm cross-section wire with a 2~mm diameter cooling channel. These are wound in their own custom made formers on the lathe and are again bound using Araldite. Coil 3 is then placed and secured over the end of the solenoid and coil 4 is secured at the end of the Zeeman slower former tube.

\begin{figure}
	\centering
		\includegraphics[width=1\linewidth]{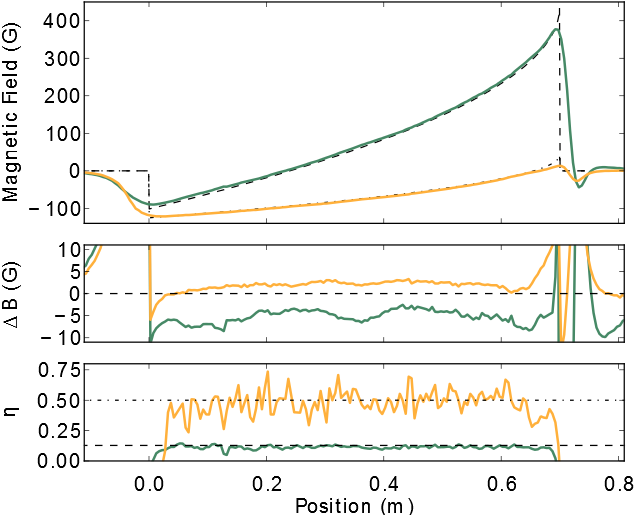}
	\caption{The top plot shows the axial magnetic field profile produced by the Zeeman slower whilst running at Yb currents (green) and Cs currents (yellow) compared with the ideal field profile for each element, shown by the dashed and dotted-dashed lines for Yb and Cs respectively. The origin is the position at which the ideal Zeeman slower field profile begins. The middle plot shows the residuals from the ideal profile along the slower. The bottom plot shows the measured design parameter, $\eta$, along the slower and the horizontal dashed (Yb) and dotted-dashed (Cs) lines show the target values of $\eta$.}
	\label{fig:ZeemanField}
\end{figure}

Prior to assembling the vacuum system, the Zeeman slower field profile is measured using an axial Hall probe (see figure~\ref{fig:ZeemanField}). This is an iterative process, with each layer pair of coils 1 and 2 being tested and compared to the ideal slower field profile. This allows turns to be added and removed from each layer in order to fine tune the field profile to that required. The top plot in figure~\ref{fig:ZeemanField} shows the final measured field profile. The field profile for Cs deviates from the designed profile by no more than 2~G along the tapered part of the Zeeman slower~\cite{Hopkins2015}. The profile for Yb is offset from the designed field profile by approximately -5~G, this is subsequently corrected by increasing the solenoid current to shift the field profile along the entire length of the slower. The parameter $\eta$ is proportional to the product of the local field and its gradient: $B\frac{dB}{dz}$.  Hence we also show in figure~\ref{fig:ZeemanField} the measured variation of $\eta$ along the length of the Zeeman slower and see that, despite the local field deviations, it always remains safely less than 1 (as it must do for success) and close to the design targets of 0.128 and 0.5 for Yb and Cs respectively.


\section{Laser systems}
\label{sec:lasers}

For both Cs and Yb we use several sources of resonant laser light in order to decelerate, cool and detect the atoms inside the apparatus. With the exception of the Yb Zeeman slowing light, where laser power is at a premium, all the required laser light is derived on a separate optical table from the main experiment.  The light is then delivered to the experiment through polarization maintaining optical fibers.

\begin{figure}
	\centering
		\includegraphics[width=1\linewidth]{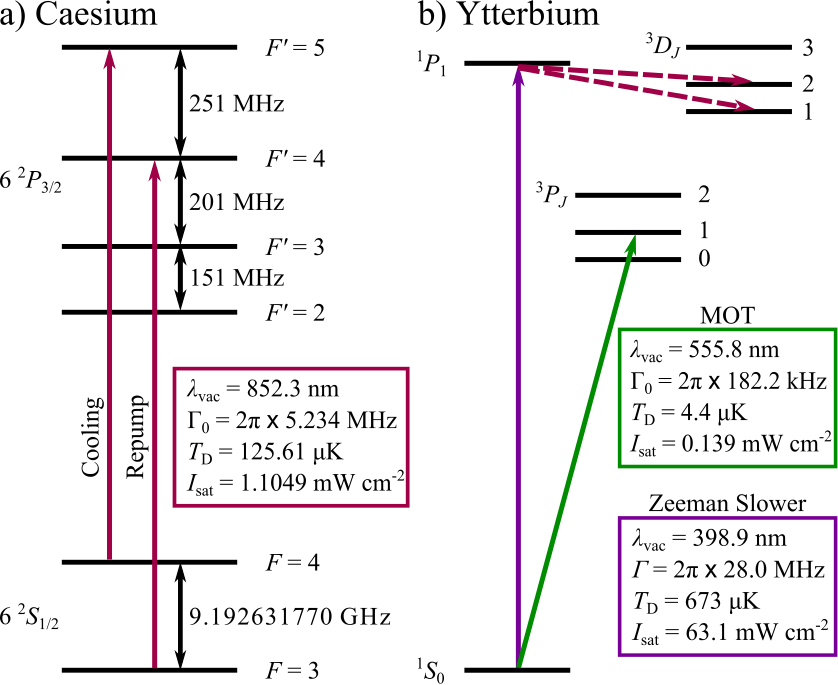}
	\caption{(a) Schematic showing the level structure of the Cs \emph{D}2 lines. Atoms are cooled on the $F=4 \rightarrow F'=5$ cycling transition. Those that fall into $F=3$ are returned to this cycle by the repump laser tuned to the $F=3 \rightarrow F'=4$ transition. The wavelength, $\lambda$ natural linewidth, $\Gamma_{0}$ Doppler temperature, $T_{\rm{D}}$, and saturation intensity, $I_{\rm{sat}}$ of the $D_{2}$ lines are indicated in the box  (b) The level structure of Yb, showing the singlet transition used for Zeeman slowing and the triplet transition used for the narrow line MOT. The weak decays from the $^{1}P_{1}$ state to the $^{3}D_{J}$ states are indicated by the dashed arrows. The wavelength, $\lambda$ natural linewidth, $\Gamma_{0}$, Doppler temperature, $T_{\rm{D}}$, and saturation intensity, $I_{\rm{sat}}$ of the blue and green transitions are shown in their respective colored boxes.}
	\label{fig:levels}
\end{figure}

Figure~\ref{fig:levels} shows the energy levels and parameters relevant for cooling both Cs and Yb. Their favorable level structures make the cooling and trapping of both species relatively straightforward, though different in character due to their differing numbers of valence electrons. Cs, with its single valence electron, has ground state hyperfine structure,  so needs light to repump atoms that have decayed into the dark $F=3$ hyperfine ground state, back into the cooling cycle. Yb has no ground state hyperfine splitting. The Yb transition from  $^{1}S_{0}$ to $^{1}P_{1}$ at 398.9~nm with a width of 28~MHz has a branching ratio from the $^{1}P_{1}$ to the $^{3}D_{2}$ and $^{3}D_{1}$ states, which removes atoms from the cooling cycle. The $^{1}S_{0}$ to $^{3}P_{1}$ transition at 555.8~nm is completely closed with a linewidth of 182.2~kHz, which is wide enough for direct loading of a MOT. Use of this transition has the advantage of a low Doppler temperature of 4.4~$\mu$K, thus circumventing the lack of sub-Doppler cooling mechanisms in Yb. For these reasons we have chosen to use the 398.9~nm transition for Zeeman slowing, exploiting the large  $a_{\rm{max}}$ on this transition. The slowed atoms are then directly loaded into the MOT using the 555.8~nm transition.


\subsection{Cesium}

The cooling light for Cs (figure~\ref{fig:levels}(a)) is provided by a Toptica DL 100 Pro laser (100~mW) which seeds a Toptica BoosTA tapered amplifier capable of 1 W output.  The repump light is provided by a Toptica DL Pro (80~mW).  Acousto-optical modulators (AOMs) shift the frequency of the MOT, imaging, and Zeeman slowing beams to their required detunings. Each AOM generating light for use on the main table is backed up by a shutter in order to prevent any unwanted resonant light reaching the atoms during later stages of the experiment. The lasers are frequency stabilized using modulation transfer~\cite{McCarron2008} and frequency modulation~\cite{Bjorklund1980} spectroscopy for the cooling and repump lasers respectively.

The Zeeman slower has ($6.0 \pm 0.1$)~mW of light on the cooling transition, set to a detuning of -59~MHz from the $F=4 \rightarrow F'=5$ transition. This is sufficiently large to avoid the Zeeman beam perturbing the MOT, but still small enough to avoid a high level of off-resonant pumping losses via the $F'=4$ level lying 251~MHz below $F'=5$. The ($2.8 \pm 0.1$)~mW  of repump light optimizes at a detuning of -50~MHz. This corresponds to the Doppler shift of Cs atoms in the zero-crossing of the Zeeman slower field, where all $m_{F}$ states are degenerate and hence effectively pumped.  The Zeeman slower light is combined with the Yb Zeeman slower beam on a dichroic mirror (Thorlabs DMSP805L), and focused down the beam line to come to a waist of ($89 \pm 1$)~$\mu$m at a distance ($2.09 \pm 0.01$)~m from the Zeeman slower viewport. Simulations show that the performance of the slower is not strongly dependent on this focusing.


\subsection{Ytterbium: 399 nm}

In order to provide sufficient slowing force, the Zeeman slower for Yb is operated on the broad 398.9~nm transition (figure~\ref{fig:levels}(b)). The light for this beam is provided by two laser diodes (Nichia NDV4314) in a master-slave configuration. The master laser is an ECDL in a Littrow configuration using a 3600~lines\,mm$^{-1}$ grating (Thorlabs GH13-36U). It is is used to injection lock a slave laser, which then produces up to 70~mW of light for the Zeeman slowing beam itself. The Zeeman slower light is focused down the vacuum system with a waist of ($307 \pm 8$)~$\mu$m at ($1.93 \pm 0.01$)~m from the Zeeman slower viewport, though simulations show that this is not critical.

A key parameter of this Zeeman slower is the detuning from resonance. Because the laser beam passes through the MOT location, and the 398.9~nm transition is so broad (leading to a high scattering rate and force on the atoms), our slower is designed to be operated 609~MHz red-detuned from resonance (for a stationary atom).  In order to obtain this detuning, a small portion of the light from the master laser is split off and fiber coupled over to the laser table. Here it is double passed through a 200~MHz AOM (Gooch and Housego, M220-4A-GH11), before being split on a polarizing beam splitter cube (PBS). Both of the output beams from the PBS are then double-passed through 100 MHz AOMs (Isomet 1206C-833). The first of these beams is used for locking the laser to the atomic resonance. The second beam is fiber coupled back to the experiment table, giving 25 $\mu$W of resonant probe light for absorption imaging. We note that although the slowing light is $\sim22$ $\Gamma$ away from resonance, it is still seen to have a significant pushing effect on the Yb MOT.

The laser  is locked using a collimated atomic beam, issuing from  a capillary array identical to that in the dual species oven. This source operates at 470\,$^{\circ}$C and produces a beam which passes through two six-way crosses. The blue spectroscopy light is then passed through the second of these crosses, perpendicular to the atomic flux. A high numerical aperture lens located on the top viewport collects the fluorescence and focuses it onto a photodiode. A concave mirror located on the bottom viewport further improves light collection.  The photodiode current is converted to a voltage through a 500~M$\Omega$ T circuit \cite{BurrBrown}, and then amplified by a low-noise amplifier. The circuit is housed within a Faraday cage. The AOM frequency is dithered at a rate of 2.5~kHz. A lock-in amplifier is used to demodulate the fluorescence peak and produce an error function in order to lock the master laser -609~MHz from resonance. In this scheme only 25~$\mu$W of light is required to lock the laser.

\subsection{Ytterbium: 556 nm}

The 556\,nm light for the narrow $^{1}S_{0}$ to $^{3}P_{1}$ transition (figure~\ref{fig:levels}(b)) is provided by a doubled fiber laser operating at 1111.6~nm (MenloSystems `Orange One'). The 556~nm light is split on a PBS and each beam passes once through a 200~MHz AOM (Gooch and Housego, 46200-0.3-LTD). The majority of the available light is in a single branch of this set up. This light is split a further two times using a combination of half waveplates and PBS cubes to make three beams. These are then coupled into polarization maintaining fibers for use as the laser cooling beams.  A mechanical shutter is used to back up the AOM and prevent any undesired resonant light reaching the experiment when the AOM is turned off. 

\begin{figure}
	\centering
		\includegraphics[width=0.9\linewidth]{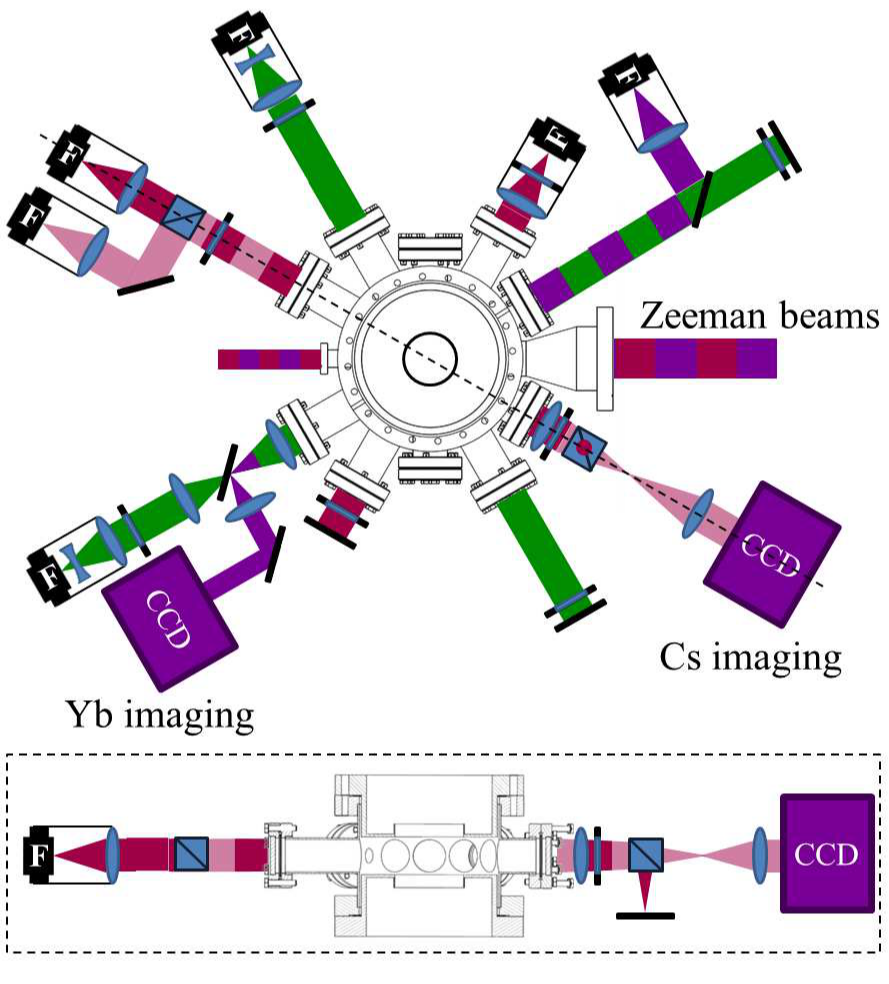}
	\caption{The top schematic shows the paths though the science chamber of the MOT, Zeeman and imaging beams for each species. All of the MOT and imaging light is delivered to the experiment using optical fibers (F). The 398.9~nm light is shown in purple, 555.8~nm light in green, and 852.3~nm light in red. The imaging light for Cs is overlapped with one of the MOT beams using a PBS cube and is shown in pink. A $\lambda/4$ waveplate, lens, and PBS separate the Cs imaging and MOT light on the opposite side of the experiment. The 398.9~nm imaging beam is overlapped with a 555.8~nm MOT beam using dichroic mirrors to combine and split the light. The dashed black line shows the plane of the cross section in the bottom schematic, which clearly shows the different paths of the Cs MOT and imaging beam. The vertical MOT beams propagate into and out of the page in the main figure though the central re-entrant viewport.}
	\label{fig:optics}
\end{figure}

The light from the second of the two AOMs is used for spectroscopy, with the difference of the two AOM frequencies giving the MOT detuning. The first-order diffracted light from this AOM is passed through the first six-way cross of the atomic beam. Similarly to the blue light, the green laser is locked by dithering the AOM frequency. A lock in amplifier is then used to derive an error signal from the modulated fluorescence.


\section{Dual species magneto-optical trapping}
\begin{figure*}[t]
	\centering
		\includegraphics[width=1\linewidth]{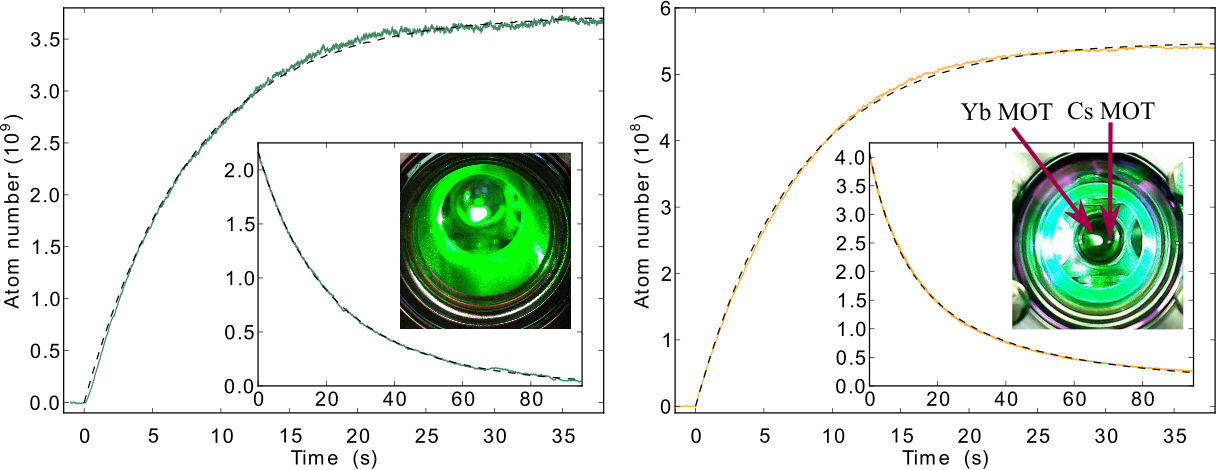}
	\caption{Loading curves for the Yb (left) and Cs (right) MOTs. The Yb MOT loads to a steady state of $3.7\times 10^{9}$ atoms when the total beam intensity is 273~$I_{\rm{sat}}$ and the detuning is -5.15~MHz. The loading rate is $6.0\times 10^{8}$ s$^{-1}$. The Cs MOT loads to a steady state of $5.5\times 10^{8}$ atoms with a total beam intensity of 12.5~$I_{\rm{sat}}$ at a detuning of -5.6~MHz. The loading rate is $7.5\times 10^{7}$~s$^{-1}$. The loading rates are extracted from a fit to equation~\ref{eq:Load}. The insets to the graph show the decay from the MOTs when the source of atoms and Zeeman slower light are turned off. A single exponential fit to the Yb curve gives a lifetime of $28.50\pm0.02$~s. For the Cs MOT a two-component exponential fit is required, giving long and short lifetimes of $40.04\pm0.02$~s and $17.44\pm0.03$~s respectively. Also shown are images of the MOTs. The Cs MOT is shown next to an Yb MOT for reference, illustrating the need for a `push' beam to overlap the two MOTs.}
	\label{fig:LoadingCurve}
\end{figure*}

The MOT is the normal starting point for ultracold atom experiments, because it allows a large number of atoms to be trapped and cooled using the scattering of resonant photons~\cite{Raab1987}. The setup in this experiment uses the standard six-beam configuration. For both Cs and Yb the MOT beams are delivered to the experiment using three fiber coupled cage systems as shown in figure~\ref{fig:optics}. The Cs MOT beams have a 1/e$^{2}$ diameter of $18.0 \pm 0.1$~mm with a power of $17.6 \pm 0.2$~mW per beam.  The amount of repump power available is $13.5 \pm 0.1$~mW and the majority of this is delivered in just one axis of the MOT beams. The Yb cage systems collimate the MOT beams to a 1/e$^{2}$ diameter of $24.5 \pm 0.2$~mm. Each MOT beam can have up to $15.0\pm 0.1$~mW of power, controlled using the AOM located before the fiber. For Cs these cages incorporate a quarter waveplate to provide the correct circular polarization for the beam. In the case of Yb, where the MOT beams are larger, a waveplate external to the cage is used to set the polarization. Each of the MOT beams in the horizontal plane passes through a pair of viewports that have been anti-reflection coated for that particular wavelength. On the downstream viewports a combination of a quarter waveplate and a mirror is used to retro-reflect the MOT beams. In the vertical direction, the Cs and Yb MOT beams are combined underneath the science chamber using a dichroic mirror, before co-propagating through the re-entrant viewports. A second dichroic mirror is used above the chamber to split the two wavelengths prior to retro-reflection.

\subsection{Ytterbium MOT}
We are able to load $3.7\times10^{9}$ Yb atoms into the MOT in 40~s, as can be seen in figure~\ref{fig:LoadingCurve}. We model this loading curve as

\begin{equation} \label{eq:Load} N = \frac{R}{\gamma}\left[ 1-\exp{\left(-\gamma t\right)}\right],\end{equation}
 where $N$ is the atom number, $R$ is the loading rate, $\gamma$ is the loss rate, and $t$ is the time elapsed. From the fit we find $R=6.0\times 10^{8}$~s$^{-1}$. We are able to achieve such a high loading rate through fine tuning of our Zeeman slower~\cite{Hopkins2015} and using the full 15.0~mW of power available per MOT beam, with a detuning of -5.15~MHz. The axial magnetic field gradient at the MOT location is 3.1~G\,cm$^{-1}$. Unlike some other experiments~\cite{Kuwamoto1999,Hansen2013}, we find that it is not necessary to add frequency side-bands to our MOT beams to achieve optimal loading rates; we find that power broadening alone is enough to bring the MOT capture velocity up to the Zeeman slower release velocity. The inset to the Yb loading curve in figure~\ref{fig:LoadingCurve} shows the decay of the MOT fluorescence upon switching off the Zeeman slower light and closing the atomic beam rotary shutter. A simple exponential decay fit to this data gives a vacuum lifetime for Yb of $28.50\pm0.02$~s.

\begin{figure}
	\centering
		\includegraphics[width=0.99\linewidth, clip=true]{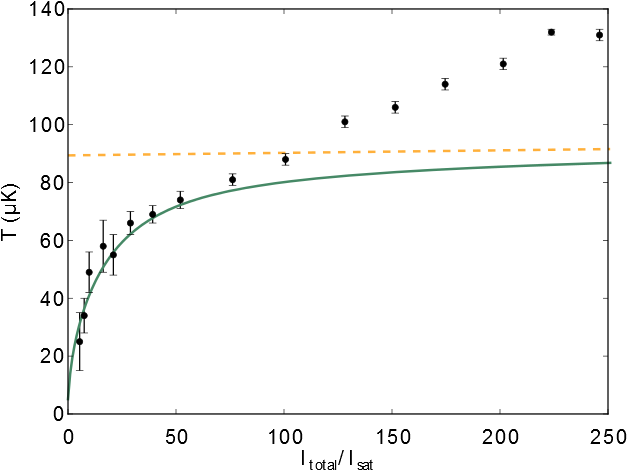}
	\caption{The temperature of the Yb MOT as a function of the total MOT beam intensity. The dashed orange curve shows the prediction of simple Doppler theory. The green curve shows the modified Doppler theory incorporating gravitational sag as set out in equations~\ref{eq:Temp} and \ref{eq:DeltaEff}}
	\label{fig:YbIntensity}
\end{figure}

\begin{figure}
	\centering
		\includegraphics[width=0.99\linewidth, clip=true]{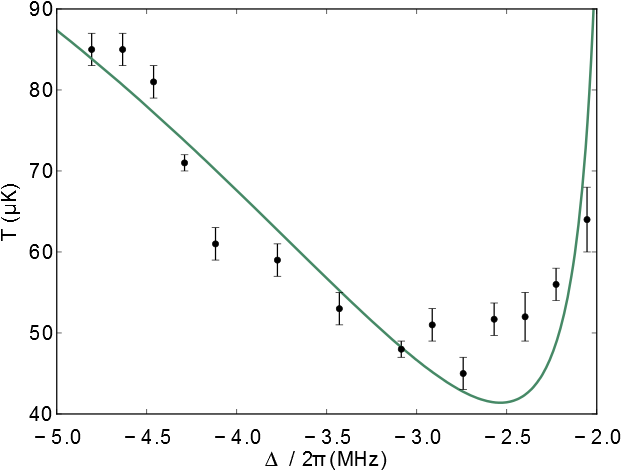}
	\caption{The temperature of the Yb MOT as a function of the red detuning from the $^{3}P_{1}$ transition at a fixed total beam intensity of 57.6~$I/I_{\rm{sat}}$. The green curve shows a fit to the modified Doppler theory incorporating gravitational sag for this intensity. The fitting parameters are a vertical scaling and a detuning offset.}
	\label{fig:YbDetuning}
\end{figure}
 
In order to calibrate the fluorescence measurements in figure \ref{fig:LoadingCurve} and to measure the temperature of the MOTs, we have implemented the standard time of flight absorption imaging technique.  The optical set up for this can be seen in figure~\ref{fig:optics}. The Yb MOT is imaged on the 398.9~nm  transition to allow easy separation from the 555.8~nm MOT light. A dichroic mirror and an interference filter mounted on the camera are sufficient to prevent charge being built up on the CCD by leaked MOT light and MOT fluorescence. The imaging sequence takes three exposures that are each 30~$\mu$s long: one with the atoms and probe light, one with the probe light alone, and a final image of just the background light. These three images are processed to give the optical depth across the cloud of atoms, from which the atom number density is inferred. The size of the cloud in time of flight expansion gives the temperature.

We find the Yb MOT temperature for the loading parameters discussed above to be $180\pm 9$~$\mu$K. There is no sub-Doppler cooling in Yb, so we have to rely upon Doppler cooling alone. To cool the atomic cloud further we ramp down the power of the MOT beams  and adjust their detuning. Figure~\ref{fig:YbIntensity} shows the effect that MOT beam intensity has on the cloud temperature for a fixed detuning of $\Delta = -2\pi \times 4.7$~MHz.

Considering the lack of hyperfine structure in the ground state of Yb, one would naively assume a linear dependence of the cloud temperature on the total MOT beam intensity, $I$, of the form~\cite{Lett1989}
\begin{equation} T = \frac{\hbar\Gamma_{0}}{8k_{\rm{B}}|\Delta|}\left[ 1+I/I_{\rm{sat}}+\left(\frac{2\Delta}{\Gamma_{0}}\right)^{2}\right]. \label{eq:Temp} \end{equation}
 
This dependence is shown by the dashed orange line in figure~\ref{fig:YbIntensity}, and clearly does not fit the data. At the lower intensities the data shown in figure~\ref{fig:YbIntensity} show more cooling than the predictions of this simple Doppler theory. These data are in the regime where $\Delta$ is greater than the power broadened linewidth, $\Gamma=\Gamma_{0}\sqrt{1+(I/I_{\rm{sat}})}$, and on the narrow $^{1}S_{0}$ to $^{3}P_{1}$ transition the force exerted by the MOT beams becomes comparable to the force due to gravity at low beam intensities. As a result, when the MOT beam intensity is reduced, the cloud drops to the  position, $z_{0}$, where the MOT force balances gravity. The atoms are then Zeeman shifted to an effective detuning of
\begin{equation} \Delta_{\rm{eff}}=\Delta + \frac{\mu}{\hbar}\frac{dB_{z}}{dz}z_{0}, \label{eq:DeltaEff}\end{equation}
where $\mu$ is the magnetic moment of the $^{3}P_{1}$ state and $\frac{dB_{z}}{dz}$ is the axial (vertical) field gradient of the MOT coils. By measuring $z_{0}$ at different MOT beam intensities we can substitute this effective detuning into equation~\ref{eq:Temp} to obtain the green curve in figure~\ref{fig:YbIntensity}. At low intensities this modified Doppler theory fits the  MOT temperature well, however at high intensities the temperatures are higher than predicted. This effect has been seen in other alkaline-earth MOTs~\cite{Maruyama2003,Xu2003} and various theories have been proposed to explain its origin, such as coherences between the excited state sublevels~\cite{Choi2008} and transverse intensity fluctuations of the MOT beams~\cite{Chaneliere2005}. The increased temperature might also be due to multiple scattering inside an optically thick cloud of atoms. We have not yet investigated the cause in detail.

Figure~\ref{fig:YbDetuning} shows the effect of detuning on the temperature of the Yb atoms, with a constant total MOT $I/I_{\rm{sat}}$ of 56.7. The green curve shows a fit to the modified Doppler theory taking the gravitational sag into account. The fitting parameters are a vertical scaling of the temperature by $\times 1.25$ and a detuning offset by 1.8~MHz.  We believe the detuning offset is caused by an offset in the locking system.  Using detuning and intensity ramps, the lowest Yb temperature obtained using these methods is $22\pm 5$~$\mu$K, a suitable temperature for transferring the atoms directly to a crossed optical dipole trap.

\subsection{Cesium MOT}

The Cs MOT can typically be loaded with $5.5\times 10^{8}$ atoms in 40~s (figure \ref{fig:LoadingCurve}), with a corresponding loading rate of $7.5\times 10^{7}$~s$^{-1}$, limited by the modest Cs oven temperature. This is achieved using the full 17.6~mW of cooling power per MOT beam, a field gradient of 8.53~G\,cm$^{-1}$, and a corresponding cooling light detuning of -5.6~MHz. The fluorescence decay curve for Cs (inset in figure~\ref{fig:LoadingCurve}) exhibits two decay rates. The Cs MOT decays quickly at first because of two-body losses, but these become negligible as the MOT density decreases. A slower loss rate remains due to collisions with background gas. The lifetime due to two-body loss is found to be $17.44\pm 0.03$~s, and that for background collisions is $20.04\pm 0.02$~s.

The Cs absorption imaging system shown in figure~\ref{fig:optics} requires the camera to be fitted with a large area shutter. This is to prevent the build up of charge on the CCD resulting from the intense MOT beams and MOT fluorescence. The opening time of this shutter limits our minimum time of flight to 11~ms.

\begin{figure}
	\centering
		\includegraphics[width=1\linewidth]{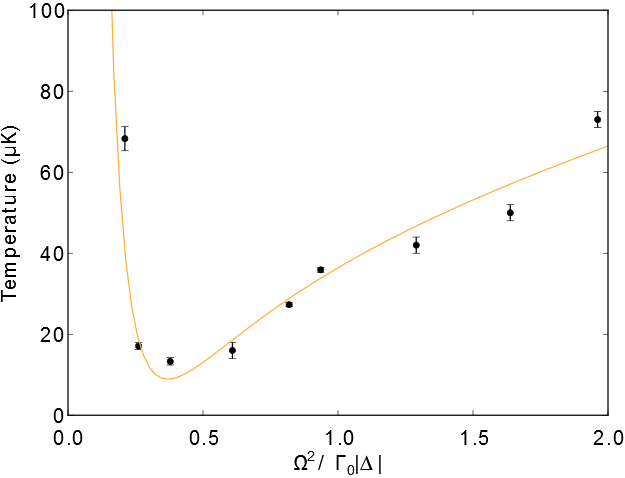}
	\caption{The temperature of the Cs optical molasses as a function of the dimensionless light shift parameter. The orange curve is a guide to the eye. The lowest temperature achieved is 13.3$\pm0.9$~$\mu$K}
	\label{fig:molasses}
\end{figure}
Figure~\ref{fig:molasses} shows the result of an experiment to measure the Cs molasses temperature. In this experiment the Cs MOT is loaded, and then the magnetic field gradient is turned off, while the Rabi frequency, $\Omega$, and the detuning, $\Delta$, of the Cs light are set to new values. After 15~ms of optical molasses, absorption imaging determines the temperature. The horizontal axis of figure~\ref{fig:molasses} is the dimensionless light-shift parameter, $\Omega^{2}/ \Gamma_{0} |\Delta | $. The temperature initially falls with decreasing light shift parameter, as expected for sub-Doppler cooling, then it rises again when the detuning and intensity imbalances are too large for sub-Doppler cooling to work effectively \cite{Castin1991}. The lowest temperature obtained of the Cs molasses is 13.3$\pm0.9$~$\mu$K.

\subsection{Dual species MOT}

This apparatus allows us to load both elements into a dual species MOT sequentially. The 398.9~nm Yb Zeeman slower photons are sufficiently energetic to ionize the trapped Cs atoms. For this reason, we load the Yb first, then the Zeeman slower field is switched to the Cs settings and, at the same time, the bias coils on the science chamber are adjusted to keep the Yb MOT central. During this step, one shutter closes to block the Yb Zeeman slower light, whilst the Cs Zeeman slower shutter opens. The Cs is then loaded with the MOT gradient held fixed at the one used for Yb, resulting in a dual species MOT. The inset to the Cs decay curve in figure~\ref{fig:LoadingCurve} shows a photo of the dual species MOT. The Cs MOT  can be seen forming to the right of the Yb MOT. This offset in position is likely to be due to an intensity imbalance in the Cs MOT beams. In future experiments we intend to use a resonant beam to push the Cs MOT in order to overlap the two MOTs prior to loading an optical dipole trap, or to load the dipole trap sequentially.


\section{Conclusion}

We have described an apparatus capable of producing a large number of co-trapped cold Cs and Yb atoms. The system makes use of a dual species oven and Zeeman slower for efficient loading into a dual species MOT. We have outlined how slow sources of Cs and Yb can be sequentially produced by the switching of the Zeeman slower coil currents. The MOT is produced in a custom built science chamber featuring re-entrant flanges that give excellent optical access as well as the ability to utilize large magnetic bias fields to aid the search for novel Feshbach resonances. We have shown that a Yb MOT operating on the intercombination line at 556~nm can accumulate a large number of atoms using a Zeeman slower working on the 398.9~nm line. In addition, we have demonstrated that by reducing the intensity of the Yb MOT beams we are able to utilize the narrowband nature of the $^{1}S_{0} \rightarrow {}^{3}P_{1}$ transition to produce samples of Yb with temperatures as low as $22\pm 5$~$\mu$K. This is similar to the Cs molasses temperature of $13.3\pm 0.9$~$\mu$K, and is an excellent starting point for loading into an optical trap. Future work will focus on measuring the interspecies scattering length of Cs and the Yb isotopes. The apparatus will be used for a full exploration of the collisional properties of Yb-Cs mixtures, and to find a route to ground state CsYb molecules.


%
%

%

\begin{acknowledgments}
We acknowledge support from the UK Engineering and Physical Sciences Research Council (grant number  EP/I012044/1) and from the Royal Society. The data presented in this paper
are available from \url{http://dx.doi.org/10.15128/73666484s}.

\end{acknowledgments}

\bibliographystyle{aipnum4-1}

%

\end{document}